\documentclass[conference,a4paper]{IEEEtran}
\usepackage{times,amsmath,mathtools,epsfig,amsfonts,amssymb,graphicx,url,cite,colortbl,bm,upgreek,multirow,booktabs,numprint}

\usepackage{standalone}
\usepackage{tikz,pgf}
\usepackage{graphicx}
\usepackage{subcaption} 
\usepackage[T1]{fontenc}
\usepackage[utf8]{inputenc}
\usepackage{pgfplots}
\usepackage{grffile}
\usepackage{comment}
\pgfplotsset{compat=newest}
\usetikzlibrary{plotmarks}
\usetikzlibrary{arrows.meta}
\usepgfplotslibrary{patchplots}
\pgfplotsset{compat=newest}
\newlength\fwidth
\begin{document}
	\title{mmWave Sensing for Detecting Movement Through Thermoplastic Masks During Radiation Therapy Treatment}
	\author{\IEEEauthorblockN{
			Ali Kourani*, %
            Naveed A. Abbasi\textsuperscript{\dag}, %
			Syeda Narjis Fatima\textsuperscript{\ddag}, %
			Katsuyuki Haneda* and %
            Andreas F. Molisch\textsuperscript{\dag}
		}                                     
		\IEEEauthorblockA{\textit{*Aalto University School of Electrical Engineering}, Espoo, Finland \\
    \textsuperscript{\dag}\textit{Viterbi School of Engineering, University of Southern California, Los Angeles, California, USA}\\
     \textsuperscript{\ddag}\textit{Keck School of Medicine, University of Southern California, Los Angeles, California, USA}\\
    Corresponding author: Ali Kourani (email: ali.kourani@aalto.fi)}
    
	} 
	\maketitle
	\begin{abstract}
 Precision in radiation therapy relies on immobilization systems that limit patient motion. Thermoplastic masks are commonly used for this purpose, but subtle voluntary and involuntary movements such as jaw shifts, deep breathing, or eye squinting may still compromise treatment accuracy. Existing motion tracking methods are limited: optical systems require a clear line of sight and only detect surface motion, while X-ray-based tracking introduces additional ionizing radiation. This study explores the use of low-power, non-ionizing millimeter-wave (mmWave) sensing for through-mask motion detection. We characterize the RF properties of thermoplastic mask material in the 28--38 GHz range and perform motion detection using a 1 GHz bandwidth centered at 28 GHz. We use a frequency-domain system with horn antennas in a custom-built anechoic chamber to capture changes in the amplitude and phase of transmitted RF waves in response to subtle head and facial movements. These findings lay some groundwork for future real-time, through-mask motion tracking and future integration with multi-antenna systems and machine learning for error correction during radiotherapy.
\end{abstract}

	
\section{Introduction}
\label{sec:introduction}

Sub-millimeter precision is critical in continuous real-time monitoring and tracking of tumors in radiation therapy treatment for precise dose delivery to  target the tumors while minimizing exposure of organs-at-risk (OARs). Thermoplastic masks help immobilize patients in conventional image-guided radiation therapy (IGRT), however, small voluntary and involuntary movements such as jaw shifts or breathing may still affect accuracy and efficacy of the treatment \cite{Kang2011_AccuratePositioning, fiorino1994skin, li2016preliminary}. 

Although optical systems have been proposed to tackle the problem described above \cite{hoisak2018role, al2022aapm}, they are inherently insufficient as they visualize the external surface only. 
Due to the use of thermoplastic immobilization masks that are not optically transparent and only expose limited facial skin through small perforations, optical surface tracking can be degraded or may effectively track the mask surface rather than the patient’s true facial surface. Optical systems are also sensitive to practical imaging conditions such as lighting variation, imperceptible patient movements during therapy, and calibration drift, which can introduce errors during precise dose delivery. As a result, even an apparently good external match to the visible surface or mask does not guarantee accurate positioning \cite{Kitagawa2022_BenefitsHeadNeckPositioning}.
On the other hand, X-ray and Cone Beam Computed Tomography (CBCT) imaging provide accurate information about internal structures but rely on ionizing radiation making them risky for frequent monitoring of healthy tissues \cite{Shao2025_RealTimeCBCT_DREME}.

There is generally good setup reproducibility across different thermoplastic mask designs.
As target volumes shrink and dose gradients steepen in stereotactic radiosurgery (SRS) and highly conformal intensity modulated radiotherapy (IMRT) or volumetric modulated arc therapy (VMAT), even small residual setup uncertainties, together with intrafraction motion and mask deformation, become clinically significant, thereby driving the need for sub-millimeter precision and tightly controlled image-guided setup protocols during radiation therapy \cite{willner1997ct, saw2001immobilization, tsai1999non}.

This creates a need for non-invasive, radiation-free methods that can detect motion through the mask. Millimeter-wave (mmWave) sensing in the 28–38 GHz range offers a promising alternative, as it penetrates plastic and the outermost skin layer and thus might be capable of capturing subtle motion through amplitude and phase changes. 
Prior studies have shown that RF sensing can detect fine movements even in the presence of occlusions \cite{Hameed2022_PushingLimitsRemoteRFSensing}. In the radiotherapy context, Olick-Gibson et al. demonstrated the feasibility of 77-81 GHz mmWave Frequency-Modulated Continuous Wave (FMCW) radar for motion monitoring during treatment delivery, including qualitative observation of motion through a face mask \cite{olick2020feasibility}. More recently, Bressler et al. focused on setup verification, estimating absolute surface distance with the FMCW mmWave radar \cite{bressler2024millimeter}. These studies establish the feasibility of mmWave sensing in radiotherapy environments, however, do not provide a dedicated characterization of thermoplastic mask effects in the considered RF regime nor a labeled, choreographed under-mask facial motion protocol aimed at analyzing specific motion types beyond basic motion detection.

In this paper, we present controlled experiments demonstrating the capability of mmWave signals to detect small head motions, including localized facial changes, behind a thermoplastic immobilization mask. Our contribution is complementary in two key respects. First, we characterize the mask’s impact on the measured RF response in our band and setup geometry to support feasibility assessment and future system design. Second, we use choreographed motion sequences with known timing and repetition, providing ground truth for the presence and type of motion and enabling analysis of under-mask facial and jaw motion patterns rather than demonstrating some motion occurrence only. Because our long-term system concept (described in Section II) relies on separating and potentially classifying motion types, these choreographed sequences provide the ground truth labels needed for supervised learning and for objective performance evaluation under mask occlusion. While the present experiments use a non-realtime setup and focus on change detection rather than clinical motion estimation, they provide a proof of principle and a labeled foundation that can support future efforts aimed at motion-type identification and spatially selective monitoring under immobilization.


The remainder of the manuscript is organized as follows. Section~II presents the motivation and describes the mmWave-based target setup that requires preliminary work covered in this paper. Section~III demonstrates the experimental setup and methodology to test measurements of the constructed table-top anechoic chamber. Section~IV presents experimental results and discussion. Section~V re-examines the target implementation addressing its feasibility and directions for future work. Finally, Section~VI concludes and summarizes the paper.

\section{\textcolor{black}{Motivation and Target Clinical Setup}}
\begin{figure}[htb]
    \centering
    \includegraphics[width=0.40\textwidth]{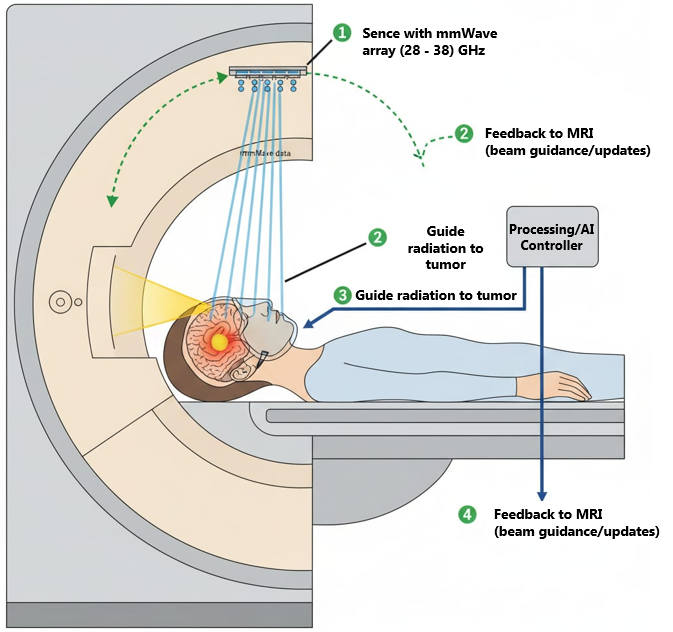}
    \caption{Envisioned clinical setup where a mmWave antenna array is integrated into the radiotherapy system to sense patient motion through the thermoplastic radiotherapy mask, with a processing unit extracting motion information and feeding real-time feedback to the treatment control system.}
    \label{fig:exp_overall}
\end{figure}

The long-term target of this work is to realize a clinical mmWave sensing system that operates inside the radiotherapy gantry and continuously monitors patient motion during treatment. Prior studies have already demonstrated in-gantry mmWave radar for radiotherapy motion monitoring and for patient setup verification, including through common immobilization materials \cite{olick2020feasibility, bressler2024millimeter}. Here we focus on a multi-view array concept aimed at spatially resolving head and face motion under a thermoplastic mask, as a complement to those single-sensor ranging approaches. As illustrated in Fig. \ref{fig:exp_overall}, a mmWave array is integrated circumferentially around the bore, in close proximity to the radiotherapy mask that immobilizes the patient. The array illuminates the head and neck region with low power signals that propagate through the thermoplastic mask, while the backscattered signals are collected across many spatially distributed elements. By exploiting beamforming and spatially distributed array locations (e.g., perpendicular to the depicted cross-section) providing different aspects, the array can achieve higher spatial resolution than a single bistatic link and can in principle localize motion within different parts of the head and face, which is essential for distinguishing global shifts from more localized jaw or facial movements. A dedicated processing and artificial intelligence (AI) unit interprets these measurements and provides motion estimates to the treatment control system in realtime, where they can support gating, beam adaptation, or table corrections that maintain the planned dose distribution.

In the envisioned clinical deployment, the mmWave system will evolve through two operational stages. In a first step, the primary task is rapid and reliable motion detection, that is, to determine on a millisecond scale whether motion above a predefined threshold has occurred while the patient is immobilized in the radiotherapy mask and to alert the treatment system accordingly. This initial capability already provides value by flagging potentially problematic motion during a fraction and prompting the operator or control software to pause or verify the treatment state. In a subsequent phase, the same sensing hardware will be used within an active feedback loop in which the measured amplitude and phase changes across the array are converted into quantitative motion estimates with spatial detail. These estimates can then be used to automatically generate system responses to a movement, such as updating beam apertures, adjusting couch position, or adapting the treatment plan, all while keeping the sensing non-invasive and free of ionizing radiation.

The present work represents a first step toward this target setup. Instead of a full array integrated into a clinical gantry, we employ a simplified frequency domain system with a pair of horn antennas in a compact anechoic chamber in order to establish basic feasibility. We first characterize the RF transparency of a clinical thermoplastic mask material in the 28--38 GHz band, demonstrating that sufficient signal energy can pass through the radiotherapy mask with limited distortion to support through-mask sensing. We then show that a single bistatic link at 28 GHz can detect subtle head and facial movements behind the immobilization device through measurable changes in the reflection/scattering coefficient. These experiments provide initial link budget and stability data, confirm that the thermoplastic mask does not preclude mmWave sensing, and supply design guidance for future multi antenna arrays that will provide the higher spatial resolution and closed loop feedback envisioned in Fig. \ref{fig:exp_overall}.


\section{Measurement Setup and methodology}

\subsubsection{Anechoic Chamber and Antenna Configuration}

The experiments were conducted using a (properly calibrated) four-port (two-port used) vector network analyzer (VNA; Keysight PNA-X N5247A) connected to transmitting and receiving horn antennas (A-Info model LB-28-15-C-KF) in a bistatic configuration inside a compact, custom-built table-top anechoic chamber lined with pyramidal RF absorber foam. As motivated in Section~II, this configuration is designed to emulate a short-range mmWave sensing link to the head region and to quantify link stability, signal-to-noise ratio (SNR), and the impact of the thermoplastic mask under controlled conditions. The device under test (DUT), i.e., the head region, with or without mask, is isolated from the rest of the body and kept stable within the test zone. In the practical clinical use case, the patient is lying down and stationary; in our custom chamber, the subject is in a seated position that provides comparable stability of the head within the illuminated region. The VNA records the complex transmission coefficient $S_{21}$ over the specified frequency spans, capturing both amplitude and phase variations of the bistatic link.

The performance of the chamber was verified through three tests. First, both antennas were placed in boresight, similar to that illustrated in Fig.~\ref{fig:exp} shaded with green but without the material sample, with a separation of 88 cm. In this configuration, the direct-path component is 42 dB stronger than any subsequent components. 
In the second verification experiment, an aluminum plate was placed at a location nearly at the subject’s head location. Together with the use of laser pointers, this ensured proper antenna orientation and the feasibility of the reflected path in the test zone, where the length of the reflected path is 1.5 m, and the leakage from the TX antenna to the RX antenna is more than 30 dB below the reflected path, which can be distinguished according to path length. The third test used a plastic mannequin with a thermoplastic mask to emulate a completely stationary subject. In this case, the round-trip length of the paths reflecting from behind the mannequin is 3.45 m, and its amplitude is more than 25 dB below the mannequin-reflected path. During these and the later experiments with a human subject, the illumination zone was completely covered with absorbers except for the subject’s head.

The antenna gain is 15 dBi, and the 3 dB beamwidth is $30^\circ$. At a distance of about 75 cm (half of the path length reflecting on the subject’s head), this beamwidth corresponds to an illumination spot of 39 cm in diameter, fully illuminating the subject’s head. As the head is completely illuminated with the $30^\circ$ beamwidth, some of the power is absorbed by the absorber frame surrounding the subject. Paths scattering on the face that do not bounce towards the receive antenna are attenuated further as a result of larger path length and the chamber absorber, and are further removed through delay gating.

\subsubsection{Thermoplastic Mask Material Characterization}

To address the first question posed in Section~II, namely whether the clinical thermoplastic mask permits sufficient mmWave transmission for through-mask sensing, the initial experiments focused on characterizing the RF transmission properties of the thermoplastic mask material. We used the antennas pointing at each other (green-shaded setup in Fig. 2), with the DUT (Blue S-Type thermoplastic mask, Klarity Health), composed of clinical-grade polycaprolactone, placed between the transmitting and receiving antennas, and no mannequin or human subject present. The antennas were separated by 67~cm ensuring that the antenna beamwidth cones illuminate the thermoplastic material and not exceed the material sample edges (residual edge-diffracted components were calculated to be attenuated by at least 20 dB). The chamber was completely enclosed with absorbers. A frequency sweep from 26.5 to 40 GHz was conducted in 10 MHz steps to measure the S-parameters and evaluate the electromagnetic transparency and frequency response of the material. 

To evaluate angle-dependent transmission behavior, the mask was incrementally rotated about its vertical axis. A 3 GHz moving window and delay gating of the main peak (shortest path) were used, and material parameters are reported for the 28--38 GHz band.
As will be quantified in Section~IV, the measured insertion loss and phase response over this band confirm that the thermoplastic mask introduces only modest attenuation and distortion, and therefore does not preclude mmWave sensing through the mask.

\subsubsection{Human-Subject Measurements and Head Fixation}

For the human-subject measurements, the measurement bandwidth is 1 GHz centered at 28 GHz, with an IF bandwidth of 50 Hz, 101 frequency points, and a sweep time of 1.7 s. Including 2.75 s of data logging, the total snapshot time per measurement is 4.45 s. These settings are chosen to provide sufficient SNR and temporal stability to quantify the magnitude of motion-induced amplitude and phase changes relative to a static baseline, even though they do not yet realize the millisecond-scale update rates envisioned for a clinical implementation.

To stabilize the subject’s head and minimize involuntary face movements during the sweep, a U-shaped wooden structure with a wide base was manufactured. The two vertical pillars in this U-shaped structure are separated by a distance that allows the subject’s head to fit between them while preventing the frame of the thermoplastic mask from passing through. The subject rests the chin on the horizontal part of the structure and relaxes the head against the two pillars, which contact the frame of the thermoplastic mask. This configuration allows the subject to support the head in the illuminated zone without relying on neck muscles, which could otherwise introduce head instability. Once the subject places the head in position, absorbers are added to cover the U-shaped structure.

Before each choreographed movement, the subject is instructed to remain still while two snapshots are collected to confirm a stable baseline. A third snapshot is then acquired when the command for movement is given, which may include the onset of motion, and a fourth snapshot is collected during the subject’s movement. In the post-processed data (for example, in Fig.~\ref{fig:stat_vs_move}), the subject’s stability is the baseline for the validity of the results, and all reported measurements include at least two stable-subject snapshots that serve as a reliable reference for the choreographed motion. These repeated baseline measurements establish the residual amplitude and phase fluctuations in the absence of intentional motion, and motion-induced changes are compared against this reference when assessing detectability.


\begin{figure}[htb]
    \centering
    \includegraphics[width=0.35\textwidth]{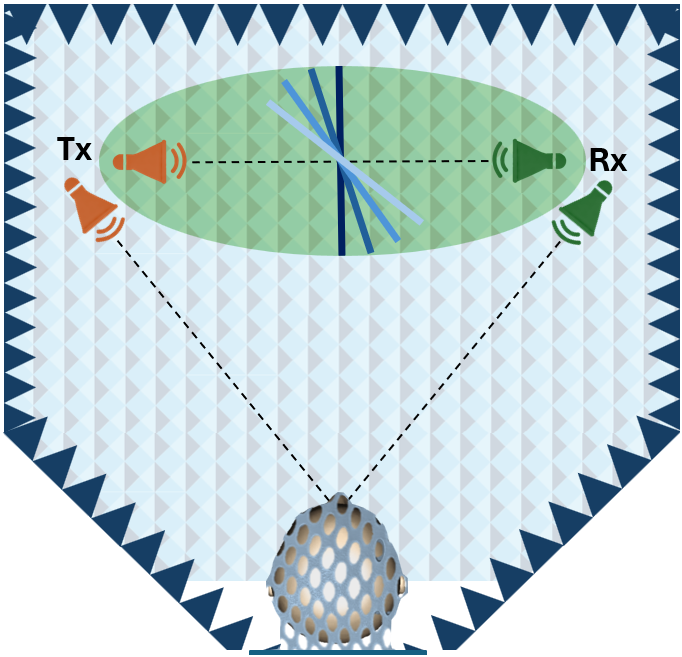}
    \caption{Experimental scenario for mannequin and human subject measurements through the thermoplastic mask, and for material characterization (highlighted in green).}
    \label{fig:exp}
\end{figure}

\section{Experimental Results}
\label{sec:res}

\subsection{Mask Material Characterization}

\begin{figure}[htb]
    \centering
    \includegraphics[width=0.35\textwidth]{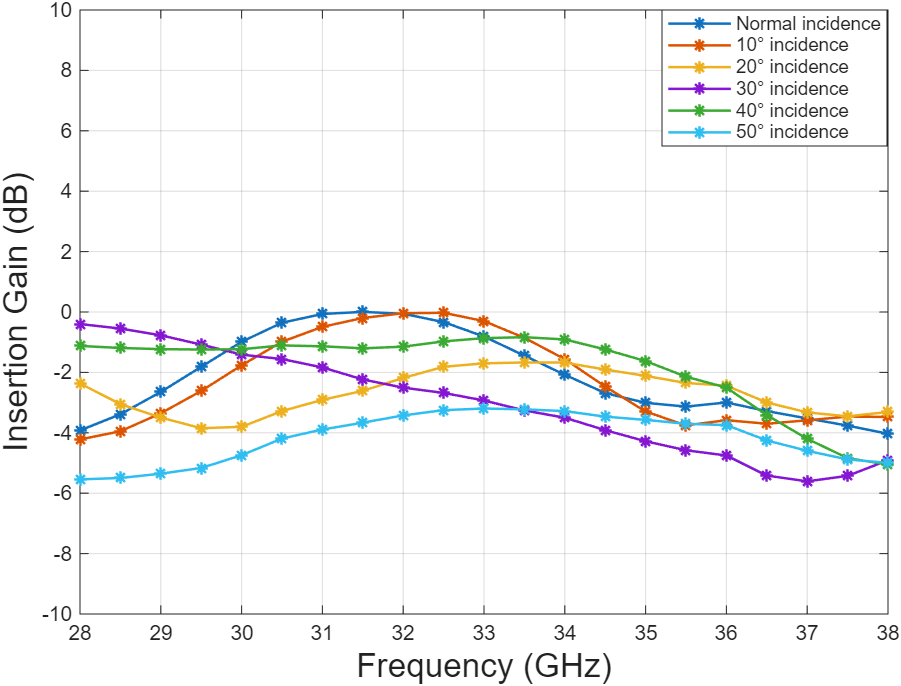}
    \caption{Results for the thermoplastic material characterization: Penetration gain over 28-38 GHz band.}
    \label{fig:Char_res}
\end{figure}

Transmission measurements in the 28–38 GHz range, shown in Fig.~\ref{fig:Char_res}, indicate low insertion loss where the highest measured value was 5.5 dB, recorded at 28 GHz at $50^\circ$ incidence, confirming that mmWave signals can propagate through the thermoplastic mask material with sufficient fidelity for reliable motion sensing applications.

\subsection{Static Mannequin Testing}

Following material characterization, a life-sized mannequin head was positioned behind the thermoplastic mask to simulate a fully stationary subject. The system recorded the transmission amplitude and phase over a 1 GHz bandwidth around 28 GHz for several minutes under no-motion conditions.

\begin{figure}[htb]
    \centering
    \includegraphics[width=0.35\textwidth]{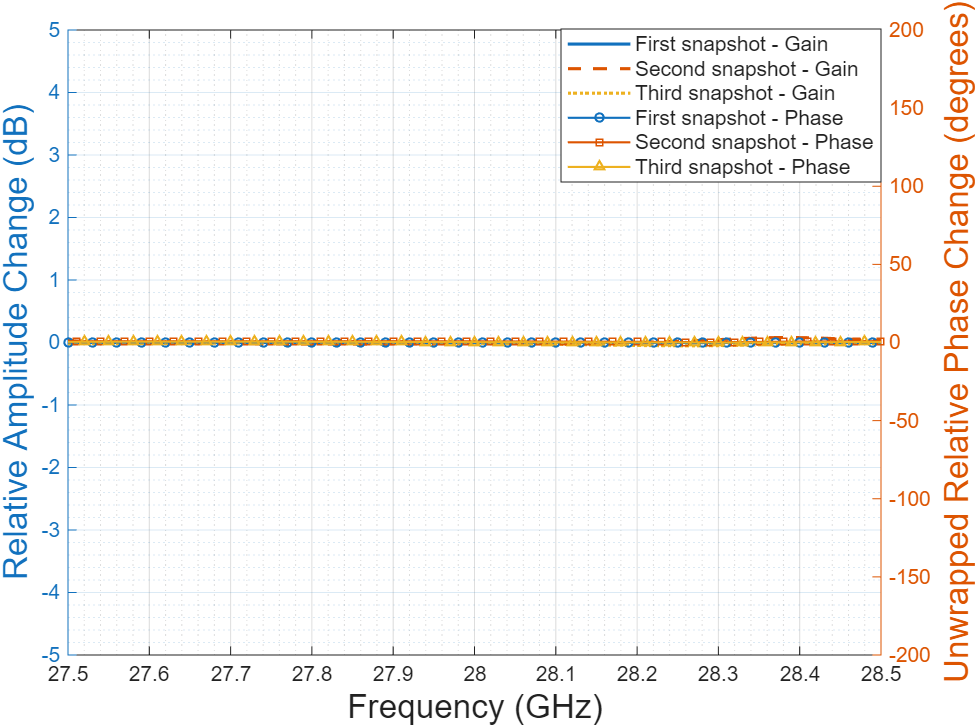}
    \caption{Unwrapped phase and amplitude variation over time when plastic mannequin is in subject's place.}
    \label{fig:manneq}
\end{figure}

Results shown in Fig~\ref{fig:manneq} demonstrate high stability in both amplitude and phase with less than 0.1 dB amplitude variation and less than $5^\circ$ phase variation, establishing a reliable static baseline. This indicated that environmental factors and system noise were well controlled, ensuring that future signal changes could be confidently attributed to subject motion.

\subsection{Human Subject Motion Detection}

In the final phase of the experiment, a healthy adult volunteer performed a series of controlled motion tests using the same setup illustrated in Fig.~\ref{fig:exp}. To replicate clinically relevant conditions, the subject remained seated behind the thermoplastic mask and performed intentional movements, including eye squinting, smiling, yawning, and slight head rotation to the right or left.

To quantify the temporal stability during the stationary phase, we evaluated the measured unwrapped phase. Over 8.9 s (two consecutive snapshots, each consisting of a 1.7 s sweep and 2.75 s data logging), the maximum unwrapped phase deviation is up to 25°, which can be attributed to involuntary human motion such as breathing. In the third snapshot, acquired during the initiation of the choreographed movement, a clearly larger phase deviation is observed, corresponding to the onset of the intended head motion. The results for the different movements and trials are reported relative to data collected while the subject was stationary. Fig.~\ref{fig:stat_vs_move} shows an example of the phase deviation between a smiling case and three preceding snapshots taken when the subject remained still.

\begin{figure}[htb]
    \centering
    \includegraphics[width=0.48\textwidth]{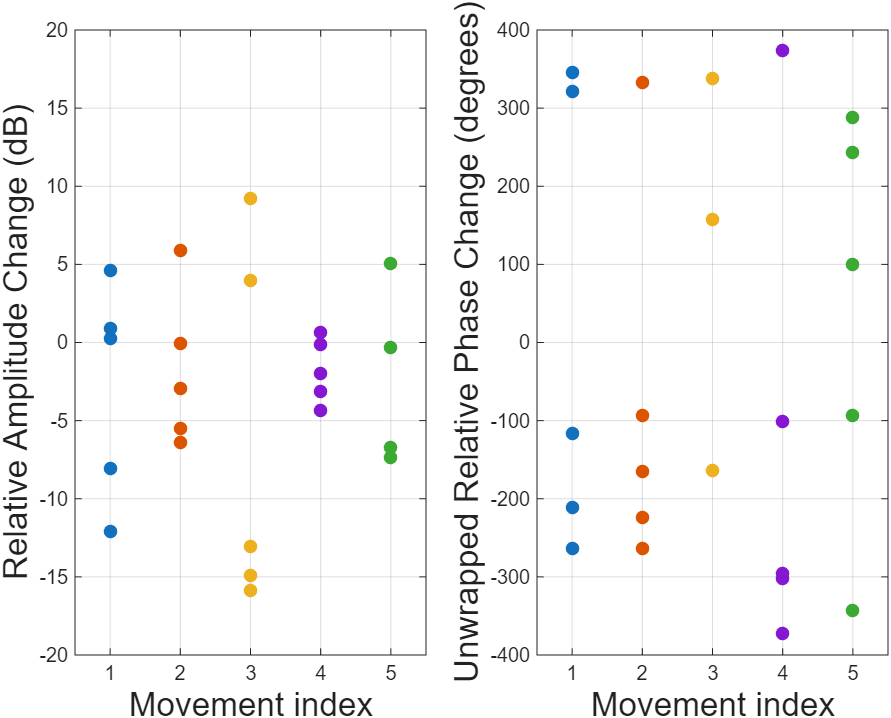}
    \caption{Amplitude and phase change at 28--GHz for each facial movement type, indexed as (1) Eye squinting, (2) Smiling, (3) Yawning, (4) Moving right, (5) Moving left.}
    \label{fig:resultss}
\end{figure}

\begin{figure}[htb]
    \centering
    \includegraphics[width=0.35\textwidth]{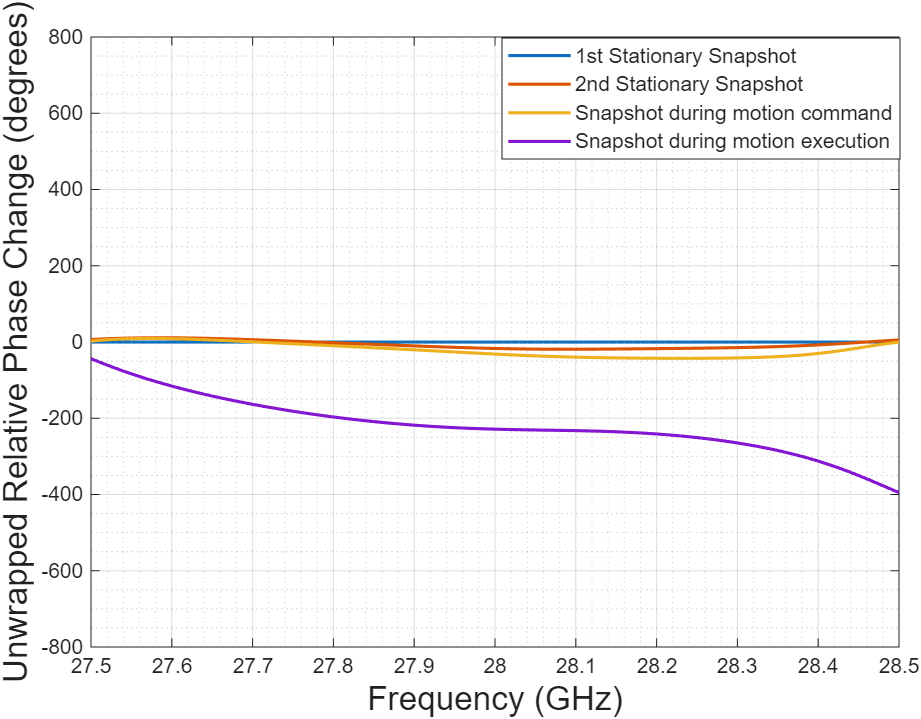}
    \caption{Example of the unwrapped phase change of the data snapshots normalized to the first snapshot of stationary subject. }
    \label{fig:stat_vs_move}
\end{figure}

The detailed results for the different facial motions are shown in Fig.~\ref{fig:resultss}. Two results are notable in these: (i) amplitude changes are often more pronounced than phase changes, though this is not always the case. For an assessment of facial movement, both quantities should be incorporated. (ii) the variations of the amplitude and phase changes between different repetitions of the same type of facial movement are larger than the difference between different types of movement. From the latter fact, we conjecture that classification of movement will require the analysis of temporal evolution, potentially via AI; this will be subject to future work. 

These measurements demonstrate that the mmWave system can distinguish static from dynamic states through the thermoplastic mask based on measurable variations in signal amplitude and phase. Although the current implementation does not yet classify specific types of motion, it confirms the feasibility of using mmWave RF sensing for non-invasive motion detection in a radiotherapy context.

\section{Conclusions}

This work demonstrates the feasibility of using millimeter-wave RF sensing to detect small facial and head movements through clinical thermoplastic masks during the setup of radiotherapy. Compared to optical systems, this approach avoids issues with mask transparency, lighting, and surface-only tracking while eliminating ionizing exposure from radiographic methods. Experimental results confirm that voluntary and involuntary motions can be reliably detected via phase and amplitude changes though discrimination between facial movement was not possible with these quantities alone. Future work will focus on integrating antenna arrays for spatial localization, adopting time-domain systems for improved sensitivity, and using machine learning to classify motion types. The ultimate goal is a non-invasive, real-time feedback system to guide and adapt radiation delivery, enhancing precision and patient safety. 
\label{sec:conclusion}

\section*{Acknowledgement}

The authors would like to thank Prof. Lijun Ma and Tathagat Pal for helpful discussions.

	\bibliographystyle{IEEEtran}
    \bibliography{ref}
\end{document}